\documentclass[reviewm]{elsarticle}

\usepackage{graphicx}
\usepackage{amssymb}
\usepackage{amsmath}
\usepackage{lineno}
\usepackage{rotating}
\usepackage{ulem}

\usepackage{color}

\newcommand{\new}[1]{#1}
\newcommand{\corrected}[1]{#1}
\newcommand{\remove}[1]{}

\newcommand{\newtwo}[1]{#1}
\newcommand{\removetwo}[1]{} 
\newcommand{\e}{\text{e}}

\begin{document}

\begin{frontmatter}

\author[eug]{Eugene d'Eon}
\author[mike]{M.M.R. Williams}
\address[eug]{8i Limited - 1/74 Cambridge Terrace - Te Aro, Wellington 6011 - New Zealand - e@8i.com}
\address[mike]{Mechanical Engineering Department -
Nuclear Engineering Group -
Imperial College of Science, Technology and Medicine -
Exhibition Road -
London, SW7 2AZ, UK}

\title{Isotropic Scattering in a Flatland Half-Space}

\begin{abstract}
  We solve the Milne, constant-source and albedo problems for isotropic scattering in a two-dimensional ``Flatland'' half-space via the Wiener-Hopf method.  The Flatland $H$-function is derived and benchmark values and some identities unique to Flatland are presented.  A number of the derivations are supported by Monte Carlo simulation.
\end{abstract}

\end{frontmatter}

\section{Intro}

  The study of linear transport theory~\cite{davison57,chandrasekhar60} in lower-dimensional spaces serves a number of purposes.  The simplicity of the one-dimensional rod model~\cite{wing62} makes it a useful tool for education~\cite{hoogenboom08a} and occasionally the starting place for exploring new general transport processes~\cite{adams89}.  Most rod model problems can be solved exactly and admit simple closed-form solutions, with diffusion encompassing the entire solution.  These properties are attractive, but distance the rod model from the complexity of three-dimemsional transport and therefore also limit its utility.

  Sandwiched between the rod model and traditional three-dimemsional scattering, two-dimensional ``Flatland'' provides a transport domain with much of the complexity of full 3D scattering, while ocassionally admitting simple closed-form solutions that have not been found in 3D (interestingly, the time-resolved Green's functions for the isotropic point source in infinite media are known exactly for 2D and 4D, but not 3D~\cite{paasschens97}).  

  Because Flatland transport research in bounded media has led to insights that improve the efficiency of 3D light transport (albeit so far participating media has not been considered~\cite{jarosz12}), we solve the classic half space problems in Flatland for isotropic scattering and investigate the form of the Flatland $H$ function and some of its numerical properties.  These derivations may aid future studies of this form in many fields.  These solutions may also directly apply to physical processes where the transport is fundamentally two-dimensional~\cite{yang09,bal2000wave,meylan06,vynck12}.

  \subsection{Related Work}
  Infinite media problems have been well studied in Flatland as well as spaces of general dimension~\cite{pearson05,kluyver06,rayleigh19,grosjean53,watson62,paasschens97,liemert11c,zoia11e,deon14} and for beams~\cite{asadzadeh08}.  Some exact solutions have been presented for bounded~\cite{liemert12c} and layered~\cite{liemert12a} media, and the singular eigenfunctions for Flatland have been derived~\cite{machida15}.  \removetwo{However, to the best of the authors' knowledge, solutions to the classic Milne and albedo problems for the half-space and the $H$-function have not been presented.}  \newtwo{Bal et al.~\cite{bal2000wave}, using an asymptotic analysis, have presented solutions to the classic Milne and albedo problems in two dimensions.  The solutions make use of the Flatland equivalent of Chandrasekhar's $H$-function, which is left as the solution to an integral equation.  We present a complementary derivation of the Milne and albedo problem solutions using the Wiener-Hopf technique.  In addition, we present new solutions and benchmark values for $H$ and provide some Monte Carlo comparisons for the albedo problem.}

  \newtwo{In the study of energy-dependent neutron transport in three-dimensional volumes with plane symmetry, Stewart et al.~\cite{stewart1966equivalence} presented a general family of solutions to the Milne problem using the method of singular eigenfunctions.  When their variable-$c$ factor $c(\xi)$ takes on the specific quantity (in our notation) $\frac{2 c}{\pi \sqrt{1-\xi^2}}$, their energy-dependent 3D solution becomes equivalent to our monoenergetic Flatland solution.  Thus, the Flatland $H$-function has, if only inadvertently, been presented long ago.
  }

 \section{General Theory}

 The Flatland one-speed transport equation can be written~\cite{asadzadeh08} as
 \begin{multline}\label{eq:transport}
   \cos \theta  \frac{\partial \phi(x,y,\theta)}{\partial x} + \sin \theta  \frac{\partial \phi(x,y,\theta)}{\partial y} + \phi(x,y,\theta) = \frac{c}{2\pi} \int_{-\pi}^\pi d\theta' \phi(x,y,\theta')+\frac{1}{2\pi} S(x,y)
 \end{multline}
 where $c=\Sigma_s / \Sigma$ and the notation is standard.  The two-dimensional analog of angular flux (or radiance) ~\cite{jarosz12} is denoted $\phi$.  If we assume that there is spatial variation in only the x-direction we find
 \begin{equation}\label{eq:transport1d}
 	\cos \theta \frac{\partial \phi(x,\theta)}{\partial x} + \phi(x,\theta) = \frac{c}{2\pi} \int_{-\pi}^\pi d\theta' \phi(x,\theta')+\frac{1}{2\pi} S(x).
 \end{equation}
 We change the angular variable in Eq.(\ref{eq:transport1d}) such that $\mu = \cos \theta$, which leads to
 \begin{equation}\label{eq:transport1dmu}
 	\left(\mu \frac{\partial}{\partial x} +1\right) \phi(x,\mu) = \frac{c}{2\pi} 2 \int_{-1}^1 \frac{d\mu'}{\sqrt{1-\mu'^2}} \phi(x,\mu')+\frac{S(x)}{2\pi}.
 \end{equation}
 For the sake of completeness we now convert Eq.(\ref{eq:transport1dmu}) to integral form for the scalar flux
 \begin{equation}\label{eq:scalarflux}
 	\phi_0(x) = 2 \int_{-1}^1 \frac{d\mu'}{\sqrt{1-\mu'^2}} \phi(x,\mu').
 \end{equation}
 Re-arranging Eq.(\ref{eq:transport1dmu}) as
 \begin{equation}\label{eq:17}
 	\frac{\partial}{\partial x} \left( \phi(x,\mu)\e^{x/\mu} \right) = \frac{1}{2 \pi \mu} \left( c \phi_0(x)+S(x)\right) \e^{x/\mu}
 \end{equation}
 and for $\mu > 0$ let us integrate from $0$ to $x$, viz:
 \begin{multline}
 	\phi(x,\mu) = \phi(0,\mu)\e^{-x/\mu} + \frac{1}{2 \pi \mu} \int_0^x dx' \left( c \phi_0(x') + S(x')\right) \e^{-(x-x')/\mu};\mu>0.
 \end{multline}
 Assuming that we have a finite slab of width $a$ we can now integate Eq.(\ref{eq:17}) from $x$ to $a$ for $\mu<0$, thus
 \begin{multline}\label{eq:19}
 	\phi(x,\mu) = \phi(a,\mu)\e^{(a-x)/\mu} - \frac{1}{2 \pi \mu} \int_x^a dx' \left( c \phi_0(x') + S(x')\right) \e^{(x'-x)/\mu};\mu<0.
 \end{multline}
 We now find the scalar flux from Eq.(\ref{eq:scalarflux}) as
 \begin{multline}\label{eq:20}
 	\phi_0(x) =\\ 2 \int_0^1 \frac{d\mu}{\sqrt{1-\mu^2}} \phi(0,\mu)\e^{-x/\mu} + \frac{1}{2\pi} \int_0^x dx' (c \phi_0(x')+S(x')) 2 \int_0^1 \frac{d\mu}{\mu\sqrt{1-\mu^2}}\e^{-(x-x')/\mu} \\
 	+ 2 \int_{-1}^0 \frac{d\mu}{\sqrt{1-\mu^2}} \phi(a,\mu)\e^{(a-x)/\mu} 
 	- \frac{1}{2\pi} \int_x^a dx' (c \phi_0(x')+S(x')) 2 \int_{-1}^0 \frac{d\mu}{\mu\sqrt{1-\mu^2}}\e^{(x'-x)/\mu}.
 \end{multline}
 Equation (\ref{eq:20}) reduces to
 \begin{multline}\label{eq:21}
 	\phi_0(x) = 2 \int_0^1 \frac{d \mu}{\sqrt{1-\mu^2}}\phi(0,\mu)\e^{-x/\mu} + 2 \int_0^1 \frac{d \mu}{\sqrt{1-\mu^2}}\phi(a,\mu)\e^{-(a-x)/\mu}\\+\frac{1}{\pi} \int_0^a dx'(c \phi_0(x')+S(x'))\int_0^1 \frac{d \mu}{\mu\sqrt{1-\mu^2}}\e^{-|x-x'|/\mu}.
 \end{multline}
 The last integral above reduces to
 \begin{equation}
 	\int_0^1 \frac{d \mu}{\mu\sqrt{1-\mu^2}}\e^{-|x-x'|/\mu} = K_0(|x-x'|).
 \end{equation}
 In Eq.(\ref{eq:21}), $\phi(0,\mu)$ and $\phi(a,\mu)$ are the incident fluxes on the faces $0$ and $a$ respectively and we may write it in general form as
 \begin{equation}
 	\phi_0(x) = I_0(x) + I_a(x) + \frac{1}{\pi} \int_0^a dx' \left( c \phi_0(x') + S(x') \right) K_0(|x-x'|).
 \end{equation}
 We now consider three classic problems in turn; Milne, albedo and constant source. \section{Milne Problem}

 The Milne problem is a special case of the above equations.  Namely, when $a=\infty$, $I_0(x) = I_a(x) = S(x) = 0$.  However, in order to solve the problem using the Wiener-Hopf technique we will return to the integro-differential Eq.(\ref{eq:transport1dmu}) with $S=0$ and the boundary condition $\phi(0,\mu)=0;\mu>0$, i.e., there is no incident current.  Also it is assumed by definition of the Milne problem that there is a continuous supply of particles from infinity that eventually leak out of the surface.  The equation to solve is
 \begin{equation}\label{eq:milne}
 	\left( \mu \frac{\partial}{\partial x} + 1\right)\phi(x,\mu) = \frac{c}{2\pi} \phi_0(x); \,\,\,\, \phi(0,\mu) = 0, \, \, \mu > 0
 \end{equation}
 We define the Laplace transform \new{with complex argument $s$} as
 \begin{equation}\label{eq:25}
 	\bar{\phi}(s,\mu) = \int_0^\infty d x \e^{-s x} \phi(x,\mu).
 \end{equation}
 Applying this to Eq.(\ref{eq:milne}) we find
 \begin{equation}
 	- \mu \phi(0,\mu) + (1+s \mu)\bar{\phi}(s,\mu) = \frac{c}{2 \pi} \bar{\phi_0}(s).
 \end{equation}
 Dividing by $1+s \mu$, multiplying by $2 / \sqrt{1-\mu^2}$ and integrating over $\mu(-1,1)$, we find after some rearrangement
 \begin{equation}\label{eq:27}
 	\left[ 1 - \frac{c}{\pi} \int_{-1}^1 \frac{d\mu}{(1+s \mu)\sqrt{1-\mu^2}}  \right]\bar{\phi_0}(s) = 2 \int_{-1}^0 d\mu \frac{\mu \phi(0,\mu)}{(1+s \mu)\sqrt{1-\mu^2}}
 \end{equation}
 where we have used the boundary condition.  The integral in the square brackets is
 \begin{equation}
 	\frac{c}{\pi} \int_{-1}^1 \frac{d\mu}{(1+s \mu)\sqrt{1-\mu^2}} = \frac{c}{\sqrt{1-s^2}}.
 \end{equation}
 Thus, we may write Eq.(\ref{eq:27}) as
 \begin{equation}\label{eq:29}
 	\left[ 1 - \frac{c}{\sqrt{1-s^2}} \right]\bar{\phi_0}(s) = 2 \int_{-1}^0 d\mu \frac{\mu \phi(0,\mu)}{(1+s \mu)\sqrt{1-\mu^2}} \equiv g(s).
 \end{equation}
 At this point we use the Wiener-Hopf method by defining
 \begin{equation}
 	V(s) = 1 - \frac{c}{\sqrt{1-s^2}},
 \end{equation}
 which has zeroes $s = \pm \sqrt{1-c^2} \equiv \pm \nu$.  We now define the function
 \begin{equation}\label{eq:31}
 	\tau(s) = \frac{s^2-1}{s^2-\nu^2} V(s) = \frac{\tau_+(s)}{\tau_-(s)}.
 \end{equation}
 The functions $\tau_\pm(s)$ are defined such that $\tau_+$ is analytic in the half-space $Re(s) < \gamma$ and $\tau_-$ in $Re(s) > -\gamma$, \new{where constant $\gamma$ is in the range $0 < \gamma < 1$}.  They are defined by
 \begin{equation}\label{eq:32}
 	\log \tau(s) = \frac{1}{2 \pi i} \int_{\gamma-i \infty}^{\gamma+i \infty} du \frac{\log \tau(u)}{u-s}-\frac{1}{2 \pi i} \int_{-\gamma-i \infty}^{-\gamma+i \infty} du \frac{\log \tau(u)}{u-s}
 \end{equation}
 and we choose the branch of the logarithm such that $\log(1) = 0$.  We may further write
 \begin{equation}\label{eq:33}
 	\log \tau(s) = \log \tau_+(s)-\log \tau_-(s)
 \end{equation}
 whence
 \begin{equation}\label{eq:34}
 	\tau(s) = \frac{\tau_+(s)}{\tau_-(s)}.
 \end{equation}
 From this we note that
 \begin{equation}\label{eq:35}
 	\tau_+(s) \tau_-(-s) = 1.
 \end{equation}
 In this section, and in the other problems, we follow the method described in Chapter 7 of~\cite{williams71}.  Thus, we can write the decomposition as
 \begin{equation}\label{eq:Vs}
   V(s) = \frac{(s^2-\nu^2)\tau_+(s)}{(s^2-1)\tau_-(s)}.
 \end{equation}
 We also have the relation (used later)
 \begin{equation}
   \tau_-(s) \tau_-(-s) = \frac{s^2-\nu^2}{(s^2-1)V(s)}.
 \end{equation}
 We may also relate $\tau_-(s)$ to the conventional $H$ function~\cite{chandrasekhar60}, namely
 \corrected{
 \begin{equation}\label{eq:38}
 	H(\mu) = \frac{(1+\mu)}{(1+\nu \mu)} \tau_-\left(\frac{1}{\mu}\right), \, \, H \left(\frac{1}{s}\right) = \frac{s+1}{s+\nu}\tau_-(s) \, \, \text{and} \, \, V(s) = \frac{1}{H(1/s)H(-1/s)}.
 \end{equation}
 }
 We now use $V(s)$ from Eq.(\ref{eq:Vs}) in Eq.(\ref{eq:29}) and re-arrange so that
 \begin{equation}\label{eq:lhsA}
 	\frac{(s^2-\nu^2)}{s+1} \frac{\bar{\phi}_0(s)}{\tau_-(s)} = \frac{s-1}{\tau_+(s)}g(s).
 \end{equation}
 As $|s|\rightarrow \infty$, the left and right hands sides of Eq.(\ref{eq:lhsA}) tend to a constant that we call $A$.  This fulfils the conditions of Liouville's theorem.  Thus, we have
 \begin{equation}\label{eq:40}
 	\bar{\phi}_0(s) = A \frac{(s+1)\tau_-(s)}{s^2-\nu^2}=A \frac{H(1/s)}{s-\nu}
 \end{equation}
 and
 \begin{equation}
 	2\frac{(s-1)}{\tau_+(s)} \int_{-1}^0 \frac{d \mu \mu \phi(0,\mu)}{(1+s \mu)\sqrt{1-\mu^2}}=A.
 \end{equation}
 Setting $s=0$ in Eq.(\ref{eq:35}) we find
 \begin{equation}
 	\frac{2}{\tau_+(0)} \int_0^1 \frac{d \mu \mu \phi(0,-\mu)}{\sqrt{1-\mu^2}} = A.
 \end{equation}
 The values of $\tau_{\pm}(0)$ are found via the method described in \cite{williams71} yielding
 \begin{equation}
 	\tau_+(0) \tau_-(0) = 1, \, \, \tau_-(0) = 1 / \sqrt{\tau(0)} = \frac{\nu}{\sqrt{1-c}}.
 \end{equation}
 Thus
 \begin{equation}
 	2 \frac{\nu}{\sqrt{1-c}} \int_0^1 \frac{d \mu \mu \phi(0,-\mu)}{\sqrt{1-\mu^2}} = A.
 \end{equation}
 Now, the current $J(x)$ at any point in the half space is defined as
 \begin{equation}
 	J(x) = -\int_{-\pi}^\pi d\theta \cos \theta \phi(x,\theta) = -2 \int_{-1}^1 d \mu \frac{\mu}{\sqrt{1-\mu^2}}\phi(x,\mu)
 \end{equation}
 from which
 \begin{equation}
 	J(0) = 2 \int_0^1 d\mu \frac{\mu}{\sqrt{1-\mu^2}}\phi(0,-\mu)
 \end{equation}
 and so
 \begin{equation}\label{eq:A}
   A = \frac{\nu}{\sqrt{1-c}} J(0).
 \end{equation}
 We also note that from Eq.(\ref{eq:19}) with $a = \infty$, that we may write
 \begin{equation}
   \phi(0,-\mu) = \frac{c}{2 \pi \mu} \bar{\phi}_0\left(\frac{1}{\mu}\right)
 \end{equation}
 i.e. the emergent angular distribution can be obtained from the Laplace transform.  From Eq.(\ref{eq:40}) \new{and Eq.(\ref{eq:A})} we find
 \begin{equation}\label{eq:48}
 	\phi(0,-\mu) = \frac{c}{2\pi} \frac{\nu}{\sqrt{1-c}}J(0) \frac{(1+\mu)\tau_-(1/\mu)}{1-\nu^2 \mu^2} = \frac{c}{2\pi} \frac{\nu}{\sqrt{1-c}}J(0) \frac{H(\mu)}{1-\nu \mu}
 \end{equation}
 where the new $H$ function is
 \begin{equation}\label{eq:49}
 	H(\mu) = \frac{(1+\mu)\tau_-(1/\mu)}{1+\nu \mu}.
 \end{equation}
 The analytic form of the function can \remove{also be obtained by the methods described in \cite{williams71}} \new{be obtained by deforming the integration contour to the imaginary axis and transforming the range of integration (see Section VII of~\cite{placzek47b})} and we find
 \begin{equation}
 	\log \tau_-(s) = - \frac{s}{\pi} \int_0^\infty \frac{dt}{s^2+t^2} \log \left[ \frac{t^2+1}{t^2+\nu^2} \left( 1- \frac{c}{\sqrt{1+t^2}}\right) \right]
 \end{equation}
 whence
 \begin{equation}\label{eq:51}
 	\tau_-\left(\frac{1}{\mu}\right) = \exp \left( \frac{-\mu}{\pi} \int_0^\infty \frac{dt}{1+\mu^2 t^2} \log \left[ \frac{t^2+1}{t^2+\nu^2} \left( 1- \frac{c}{\sqrt{1+t^2}}\right) \right] \right).
 \end{equation}
 We may simplify this integral by writing
 \begin{multline}
 	\int_0^\infty \frac{dt}{1+\mu^2 t^2} \log \left[ \frac{t^2+1}{t^2+\nu^2} \left( 1- \frac{c}{\sqrt{1+t^2}}\right) \right] = \\ \frac{1}{\mu} \int_0^\infty \log \left[ \frac{t^2+1}{t^2+\nu^2} \left( 1- \frac{c}{\sqrt{1+t^2}}\right) \right] d \left( \tan^{-1}(t \mu) \right),
 \end{multline}
 \remove{Integrating by parts, we find} \new{which upon integration by parts becomes}
 \begin{multline}
   \frac{1}{\mu}\left[ \tan^{-1}(t \mu) \log \left[ \frac{t^2+1}{t^2+\nu^2} \left( 1- \frac{c}{\sqrt{1+t^2}}\right) \right] \right]_0^\infty \\ -\frac{1}{\mu} \int_0^\infty \tan^{-1}(t \mu) \frac{d}{dt} \log \left[ \frac{t^2+1}{t^2+\nu^2} \left( 1- \frac{c}{\sqrt{1+t^2}}\right) \right]. 
 \end{multline}
 The quantity in the square brackets is zero and we find that
 \begin{equation}
   \frac{d}{dt} \log \left[ \frac{t^2+1}{t^2+\nu^2} \left( 1- \frac{c}{\sqrt{1+t^2}}\right) \right] = \frac{c t}{(t^2+1)(c+\sqrt{t^2+1})}.
 \end{equation}
 Therefore
 \begin{equation}\label{eq:55}
 	\tau_-\left(\frac{1}{\mu}\right) = \exp \left( \frac{c}{\pi} \int_0^\infty \frac{t \tan^{-1}(\mu t)}{(t^2+1)(c+\sqrt{t^2+1})} \right).
 \end{equation}
 We also have from Eq.(\ref{eq:25})
 \begin{equation}
 	\bar{\phi}_0(s) = \int_0^\infty d x \, \e^{-s x} \phi_0(x).
 \end{equation}
 Note, from this and Eq.(\ref{eq:40}) that
 \begin{equation}
   \lim_{s\to\infty} s \, \bar{\phi}_0(s) = \phi_0(0) = A = \frac{\nu}{\sqrt{1-c}} J(0).
 \end{equation}

\subsection{Spatial Variation}
 
 We may obtain the spatial variation of the scalar flux by inverting the Laplace transform
 \begin{equation}
 	\phi_0(x) = A \frac{1}{2\pi i} \int_L ds \frac{\e^{sx}(s+1)\tau_-(s)}{s^2-\nu^2}.
 \end{equation}
 The integrand has two poles at $s = \pm \nu$ and, as we will see through the structure of $V(s)$, there is a branch point at $s = -1$.  The pole contribution, which we denote $\phi_{asy}(x)$, is given by
 \begin{equation}\label{eq:59}
 	\phi_{asy}(x) = \frac{A}{2 \nu} \left( (1+\nu)\tau_-(\nu) \e^{\nu x} - (1-\nu)\tau_-(-\nu)\e^{-\nu x} \right).
 \end{equation}
 We re-write this expression in the form
 \begin{equation}\label{eq:60}
 	\phi_{asy}(x) = \frac{A}{2 \nu} B \sinh \left[ \nu(x + Z_0) \right]
 \end{equation}
 where $Z_0$ is the extrapolated endpoint, i.e. the distance into the region $x < 0$ where the asymptotic flux, mathematically, goes to zero.  By expanding Eq.(\ref{eq:60}) and comparing terms with Eq.(\ref{eq:59}) we find
 \corrected{
 \begin{equation}
 	Z_0 = \frac{1}{2\nu} \log \left[ \frac{(1+\nu) \, \tau_-(\nu)}{(1-\nu) \, \tau_-(-\nu)} \right] = \frac{1}{2\nu} \log \left[ \frac{1+\nu}{1-\nu} \right] + \frac{1}{2\nu} \log \left[ \frac{\tau_-(\nu)}{\tau_-(-\nu)} \right]
 \end{equation}
 }
 and
 \begin{equation}
 	B = 2 \sqrt{1-\nu^2} [ \tau_-(\nu)\tau_-(-\nu)]^{1/2} = 2 c [ \tau_-(\nu)\tau_-(-\nu)]^{1/2} = 2 \sqrt{2} c.
 \end{equation}
 The latter reduction follows from
 \corrected{
 \begin{equation}
 	\tau_-(\nu)\tau_-(-\nu) = \frac{1}{\tau(\nu)} = \lim_{s\to \nu}\frac{s^2-\nu^2}{s^2-1}\frac{1}{V(s)} = 2.
 \end{equation}
 }
 Hence
 \begin{equation}
 	\phi_{asy}(x) = \sqrt{2} \frac{A}{\nu} c \, \sinh[\nu(x+Z_0)].
 \end{equation}
 But we also know from \cite{williams71} that
 \begin{equation}
 	\tau_-(\nu)\tau_+(\nu) = \frac{\tau_+(\nu)}{\tau_+(-\nu)} = \frac{\tau_-(\nu)}{\tau_-(-\nu)}.
 \end{equation}
 Also
 \begin{equation}
 	\frac{1}{2\nu} \log \left[ \frac{\tau_-(\nu)}{\tau_-(-\nu)} \right] = \frac{1}{2 \pi i} \int_{\gamma-i \infty}^{\gamma+i \infty} \frac{\log \tau(u)}{u^2-\nu^2}du.
 \end{equation}
 Again, using methods described in \cite{williams71}, we find
 \begin{equation}
 	\frac{1}{2 \pi i} \int_{\gamma-i \infty}^{\gamma+i \infty} \frac{\log \tau(u)}{u^2-\nu^2}du = -\frac{1}{\pi} \int_0^1 \frac{dt}{1-\nu^2 t^2} \tan^{-1}\left( \frac{c \, t}{\sqrt{1-t^2}} \right)
 \end{equation}
 and so
 \begin{equation}
 	Z_0 = \frac{1}{2\nu} \log \left[ \frac{1+\nu}{1-\nu} \right] - \frac{1}{\pi} \int_0^1 \frac{d t}{1-\nu^2 t^2} \tan^{-1}\left( \frac{c t}{\sqrt{1-t^2}} \right).
 \end{equation}
 Note that for the purely scattering case $c = 1$ we find $Z_0 = \frac{1}{2} + \frac{1}{\pi} = 0.818309886...,$  \new{ which differs from the well-known value of $0.7104...$ for transport in three dimensions.  To the best of our knowledge, this constant has not previously appeared in the literature.}  To obtain the contribution from the branch point we return to Eq.(\ref{eq:40}) and use Eq.(\ref{eq:31}) to get
 \corrected{
 \begin{equation}
   \phi_0(x) = A \frac{1}{2 \pi i} \int_L ds \frac{\e^{s x}}{(s-1) \tau_-(-s)} \frac{\sqrt{1-s^2}}{\sqrt{1-s^2}-c}.
 \end{equation}
 }
 The contour $L$ may be deformed so that it encloses the poles and hence leads to the asymptotic part of the solution, but also it is wrapped around the branch cut, which runs from $-1$ to $-\infty$.  After some algebra and noting that
 \begin{equation}
 	\sqrt{1-s^2} = \sqrt{|1-s^2|}\e^{\pm i \pi / 2}
 \end{equation}
 where $+$ refers to the upper side of the cut and $-$ to the lower side, we get
 \begin{align}
 	\phi_{trans}(x) &= -A \frac{c}{\pi} \int_0^1 d t \frac{\sqrt{1-t^2} \e^{-x/t}}{(1+t)(1-\nu^2 t^2) \tau_-(1/t)} \\ &= - \frac{\nu}{\sqrt{1-c}} J(0) \frac{c}{\pi} \int_0^1 dt \frac{\sqrt{1-t^2} \e^{-x/t}}{(1-\nu^2 t^2)(1+\nu t) H(t)}.
 \end{align}
 The complete spatial variation of the flux is
 \begin{equation}
 	\phi_0(x) = \phi_{asy}(x) + \phi_{trans}(x).
 \end{equation} \section{The Flatland H-function}

\begin{figure}
        \centering
        \includegraphics[width=\linewidth]{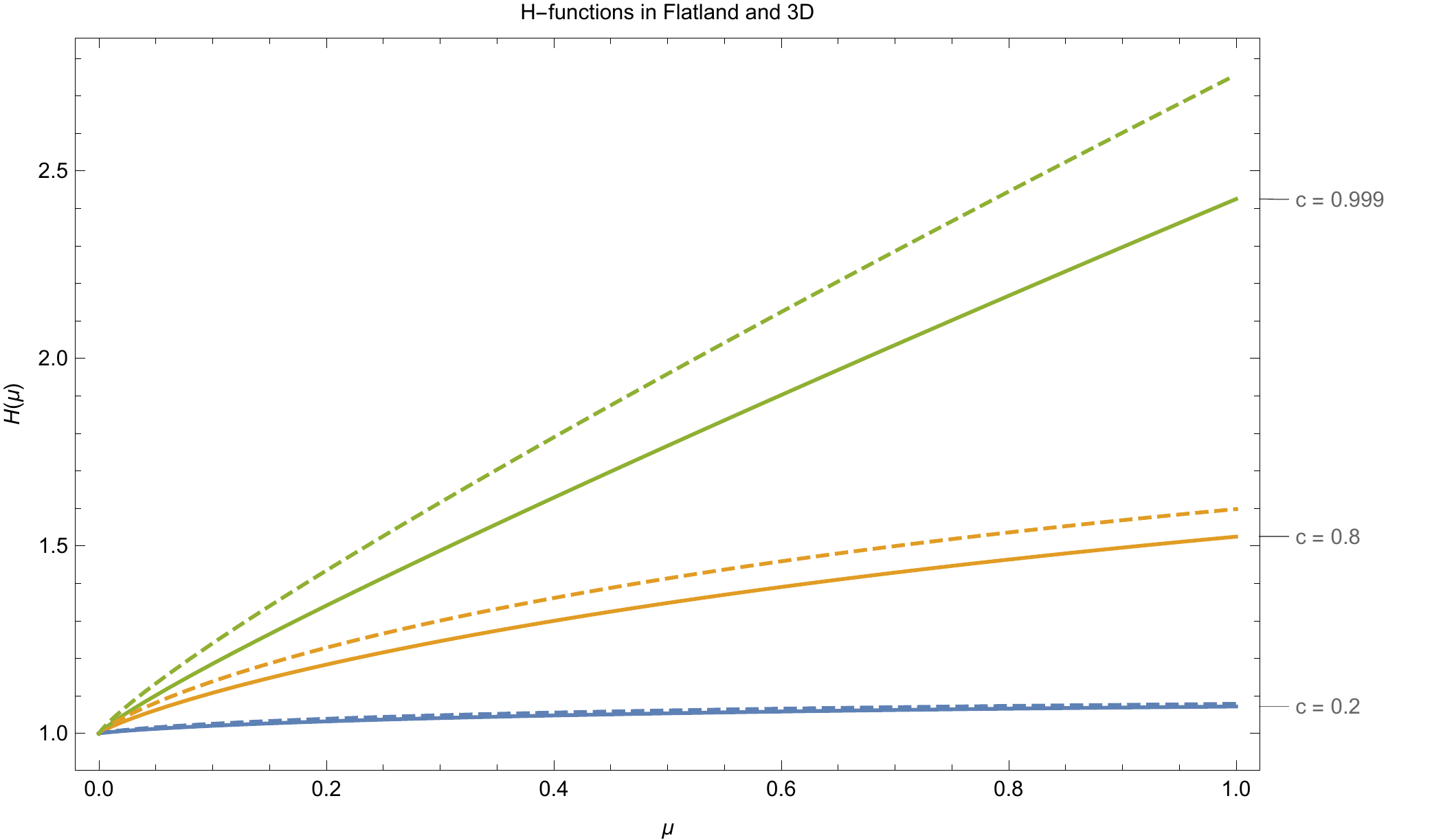}
        \caption{\label{fig:Hplot}Flatland H-function $H(\mu)$ plotted for 3 levels of absorption (continuous plots) with the standard 3D H-function shown for comparison (dashed).}
      \end{figure}

  There are some useful identities that may be obtained between the H functions.  To get these let us return to Eq.(\ref{eq:29}) and use Eq.(\ref{eq:38}) in the form
  \begin{equation}
  	\frac{1}{H(1/s)H(-1/s)} \bar{\phi}_0(s) = -2 \int_0^1 d \mu \frac{\mu \phi(0,-\mu)}{(1-s\mu)\sqrt{1-\mu^2}}.
  \end{equation}
  Inserting Eq.(\ref{eq:40}) for $\bar{\phi}_0(s)$ and Eq.(\ref{eq:48}) for $\phi(0,-\mu)$ we find
  \begin{equation}
  	\frac{1}{(s-\nu)H(-1/s)} = -\frac{c}{\pi} \int_0^1 d \mu' \frac{\mu' H(\mu')}{\sqrt{1-\mu'^2}(1-s \mu')(1-\nu \mu')}.
  \end{equation}
  But we may write
  \begin{equation}
  	\frac{\mu'}{(1-s \mu')(1-\nu \mu')} = \frac{1}{s-\nu} \left( \frac{1}{1-s \mu'} - \frac{1}{1 - \nu \mu'} \right),
  \end{equation}
  which leads to
  \begin{equation}\label{eq:75}
  	\frac{1}{H(-1/s)} = -\frac{c}{\pi} \int_0^1 d \mu' \frac{H(\mu')}{\sqrt{1-\mu'^2}(1-s \mu')} + \frac{c}{\pi} \int_0^1 d \mu' \frac{H(\mu')}{\sqrt{1-\mu'^2}(1-\nu \mu')}.
  \end{equation}
  \new{Now let $s \to \infty$ to get, with $H(0) = 1$},
  \begin{equation}
    \frac{c}{\pi} \int_0^1 d \mu' \frac{H(\mu')}{\sqrt{1-\mu'^2}(1-\nu \mu')} = 1.
  \end{equation}
  Set $s = 0$ in Eq.(\ref{eq:75}) and use $H(-\infty) = 1 / \sqrt{1-c}$ to get
  \begin{equation}
  	\frac{c}{\pi} \int_0^1 d \mu' \frac{H(\mu')}{\sqrt{1-\mu'^2}} = 1 - \sqrt{1-c}.
  \end{equation}
  Finally, setting $s = -1/\mu$, in Eq.(\ref{eq:75}) we get
  \begin{equation}
  	H(\mu) = 1 + \frac{c}{\pi} \mu H(\mu) \int_0^1 d \mu' \frac{H(\mu')}{\sqrt{1-\mu'^2}(\mu+\mu')},
  \end{equation}
  which is \newtwo{the Flatland} equivalent to \newtwo{Chandrasekhar's} equation for conventional radiative transfer.  \new{Figure~\ref{fig:Hplot} compares this Flatland H-function to the traditional 3D form for several absorption levels.  In~\ref{appendix} we provide benchmark values, new identities and additional forms for numerical evaluation of $H$.}

   \section{Constant Source Problem}

  The equation in this case is
  \begin{equation}
  	\left( \mu \frac{\partial}{\partial x} + 1 \right) \phi(x,\mu) = \frac{1}{2\pi} \left( c \phi_0(x) + S_0 \right); \, \, \, \, \phi(0,\mu) = 0, \, \mu > 0.
  \end{equation}
  Following the Wiener-Hopf procedure as described above we may write the solution for the emergent radiation as
  \begin{equation}
  	\phi(0,-\mu) = \frac{S_0}{2 \pi \sqrt{1-c}} H(\mu) 
  \end{equation}
  and the scalar intensity at the surface as
  \begin{equation}
	\phi_0(0) = \frac{S_0}{c} \left( \frac{1}{\sqrt{1-c}}-1 \right).
  \end{equation}

\section{Albedo Problem}
  The equation in this case is
  \begin{equation}
  	\left( \mu \frac{\partial}{\partial x} + 1 \right) \phi(x,\mu) = \frac{c}{2\pi} \phi_0(x); \, \, \, \, \phi(0,\mu) = \delta(\mu-\mu_0), \, \mu > 0.
  \end{equation}
  Again, following the procedure described above, we find
  \begin{equation}
  	\phi(0,\mu) = \frac{c}{2 \pi} \frac{\mu_0 H(\mu_0) H(\mu)}{\mu_0 + \mu}
  \end{equation}
  and
  \begin{equation}
  	\phi_0(0) = \phi_0(0;\mu_0) = H(\mu_0).
  \end{equation}
  Thus, the two-dimensional BRDF~\cite{jarosz12} for the isotropic Flatland half space is
  \begin{equation}
  	f_r(\theta_i,\theta_o) = \frac{c}{2 \pi} \frac{ H(\cos \theta_i) H(\cos \theta_o)}{\cos \theta_i + \cos \theta_o}.
  \end{equation}
  A Taylor series expansion about $c=0$ of $f_r$ gives the BRDFs for the singly- and doubly- scattered (and higher order) reflectances
  \begin{equation}
    f_1(\theta_i,\theta_o) = \frac{c}{2 \pi} \frac{ 1 }{\cos \theta_i + \cos \theta_o}
  \end{equation}
  and
  \begin{equation}
    f_2(\theta_i,\theta_o) = \frac{c^2 \left(\frac{\sec ^{-1}\left(\cos \left(\theta _i\right)\right)}{\sqrt{1-\sec
   ^2\left(\theta _i\right)}}+\frac{\sec ^{-1}\left(\cos \left(\theta
   _o\right)\right)}{\sqrt{1-\sec ^2\left(\theta _o\right)}}\right)}{2 \pi ^2 \left(\cos
   \left(\theta _i\right)+\cos \left(\theta _o\right)\right)}
  \end{equation}
  respectively.  The surface flux can be averaged over all incident directions to give
  \begin{equation}
  	\int_0^1 \frac{d \mu_0}{\sqrt{1-\mu_0^2}} \phi_0(0;\mu_0) = \int_0^1 \frac{d \mu_0}{\sqrt{1-\mu_0^2}} H(\mu_0) = \frac{\pi}{c}(1-\sqrt{1-c}).
  \end{equation}
  Thus, the total albedo from the half space under illumination arriving at cosine $\mu_i$ is
  \begin{equation}\label{eq:totalalbedo}
	R(c,\mu_i) = 1 - \sqrt{1-c} H(\mu_i),
  \end{equation}
  with singly-scattered contribution
  \begin{equation}\label{eq:albedoSingle}
      	R_1(c,\mu_i) = c \left(\frac{1}{2}-\frac{\sec ^{-1}\left(\mu _i\right)}{\pi  \sqrt{1-\frac{1}{\mu
   _i^2}}}\right)
      \end{equation}
  and doubly-scattered contribution
  \begin{equation}\label{eq:albedoDouble}
      	R_2(c,\mu_i) = c^2 \left(\frac{1}{8} \left( \frac{1-\mu_i}{1+\mu_i} \right)+\frac{\sec ^{-1}\left(\mu _i\right)}{2 \pi  \sqrt{1-\frac{1}{\mu
   _i^2}}}+\frac{\mu _i^2 \sec ^{-1}\left(\mu _i\right){}^2}{2 \pi ^2 \left(1-\mu
   _i^2\right)}\right).
  \end{equation}

  \begin{figure}
	      \centering
	      \includegraphics[width=\linewidth]{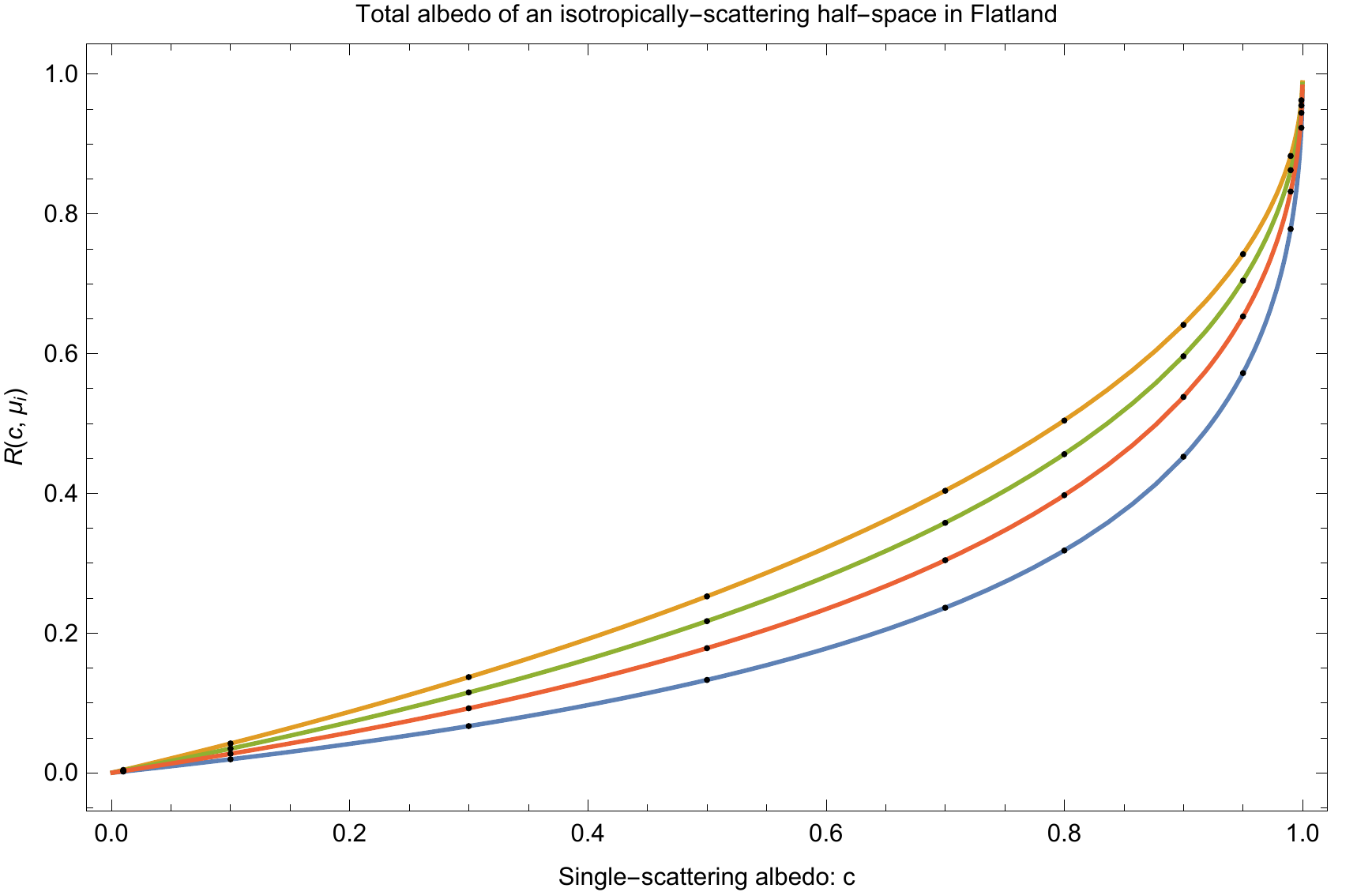}
	      \caption{Monte Carlo evaluation for the Flatland half-space albedo $R$.  Four incidence values are plotted $\mu_i \in \{ 0.1, 0.25, 0.5, 1.0 \}$, Monte Carlo is shown as dots.\label{fig:albedo}}
	    \end{figure}

	  \begin{figure}
	      \centering
	      \includegraphics[width=\linewidth]{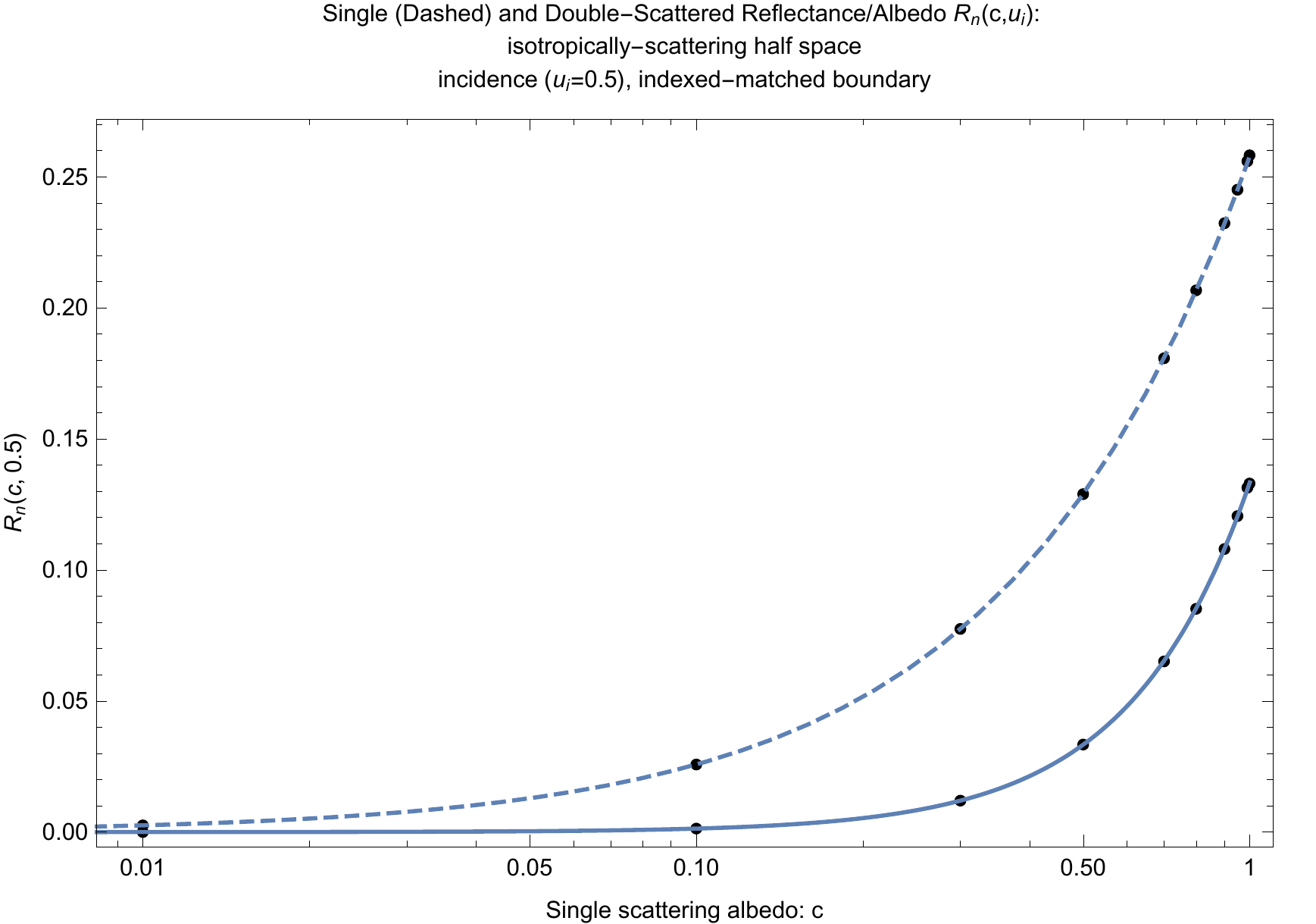}
	      \caption{Monte Carlo evaluation for Flatland half-space single-scattering albedo $R_1$ and double-scattering albedo $R_2$.\label{fig:albedoSingleDouble}}
	    \end{figure}

      \begin{figure}
        \centering
        \includegraphics[width=\linewidth]{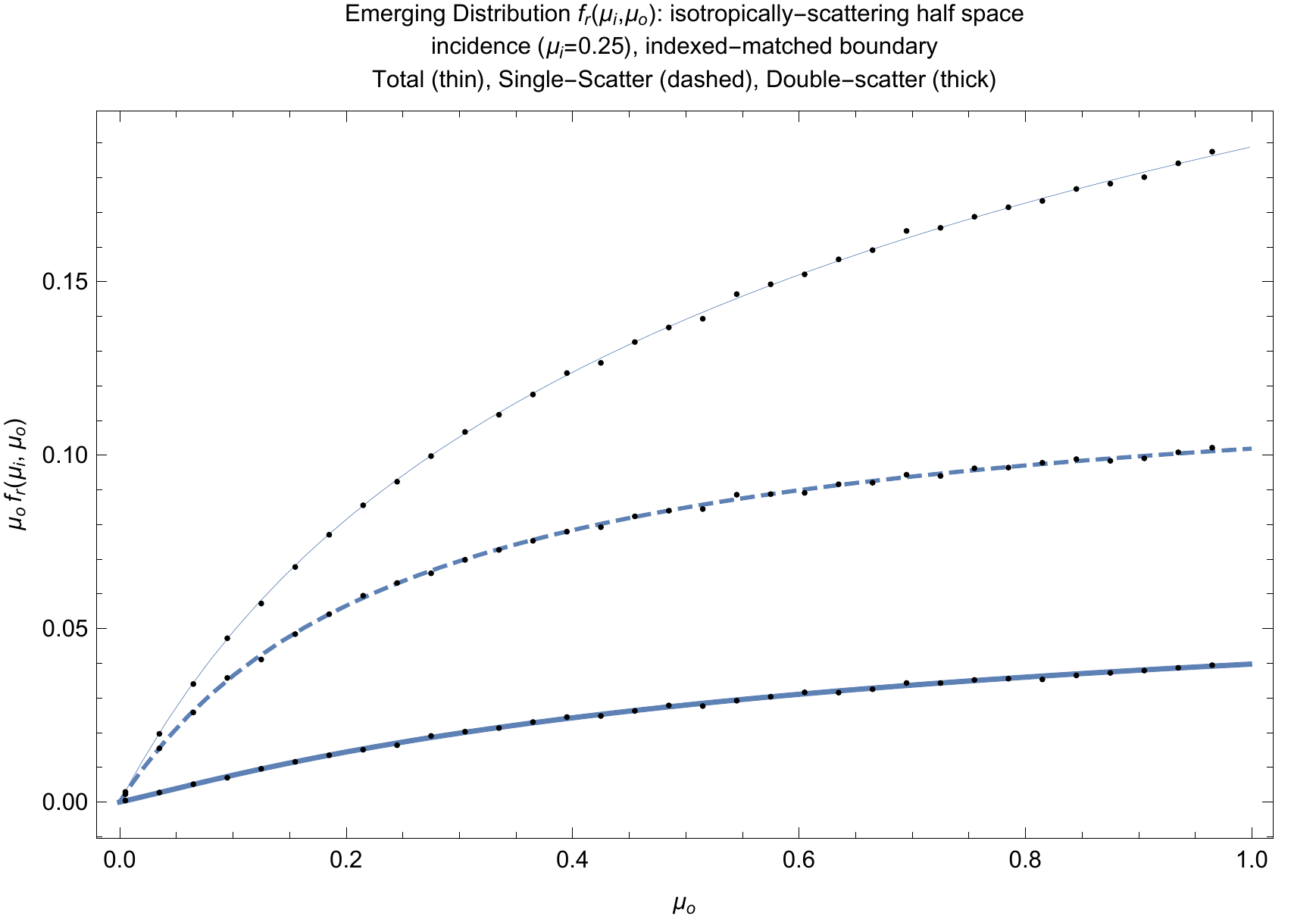}
        \caption{Monte Carlo evaluation for Flatland half-space BRDF and its single- and double-scattered portions \new{for the case of $c=0.8$}.\label{fig:brdf}}
      \end{figure}
  
 \section{Numerics}

\new{We performed Monte Carlo simulation in a two-dimensional domain to test several of the previous derivations.} 
The total albedo from the half space (Eq.(\ref{eq:totalalbedo})) is validated in Figure~\ref{fig:albedo} for four incidence angles $\mu_i$.  The singly- and doubly-scattered portions of the albedo (Eqs.(\ref{eq:albedoSingle}) and (\ref{eq:albedoDouble})) are validated in Figure~\ref{fig:albedoSingleDouble}.  The BRDF for the half space is validated in Figure~\ref{fig:brdf}. \section{Conclusion}

  We have solved the classic half-space problems of isotropic scattering for transport in a two-dimensional Flatland domain.  We found that the solutions correspond closely to Monte Carlo simulation.  A variety of forms for numerically evaluating the H-function have been provided together with benchmark values.  

  The Flatland solutions have a similar form to the familiar three-dimensional half space.  However, a new H-function identity unique to Flatland as well as analytic expressions of the H-function in terms of hypergeometric functions have been presented.

\section{Ackowledgements}
  \newtwo{We thank Norm McCormick for pointing out several important related works~\cite{stewart1966equivalence,bal2000wave}.} 
\appendix
\section{Numerical Benchmarks}\label{appendix}

  \new{In this appendix we derive additional forms of $H$ and compare their performance and accuracy in numerical integration contexts}.  \new{For the remainder of the paper we assume $0 \leq \mu \leq 1$ and $0 \leq c \leq 1$}.  Combining Eqs.(\ref{eq:49}) and (\ref{eq:51}) and simplifying the argument of the $log$, we can write
  \begin{equation}
    H(\mu) = \frac{(1+\mu)}{1+\nu \mu} \exp \left(\frac{\mu}{\pi}\int_0^{\infty } \frac{ dt}{ 1+t^2 \mu^2} \log
   \left(1+\frac{c}{\sqrt{1+t^2}}\right) \,
   \right).
  \end{equation}
  Expanding the $\log$ argument into even and odd functions of $c$
  \begin{equation}
    \log \left( 1 + \frac{c}{\sqrt{1+t^2}} \right) = \frac{1}{2} \log \left(1-\frac{c^2}{t^2+1}\right)+\tanh
   ^{-1}\left(\frac{c}{\sqrt{t^2+1}}\right)
  \end{equation}
  and noting that
  \corrected{
  \begin{equation}
    \int_0^\infty dt \frac{1}{t^2+\mu^2} \frac{1}{2} \log \left(1-\frac{c^2}{t^2+1}\right) = -\frac{\pi}{2 \mu}  \log \left(\frac{1+\mu}{1+\nu \mu}\right)
  \end{equation}
  }
  allows two additional forms of H,
  \begin{equation}
    H(\mu) = \exp \left(\frac{-\mu}{\pi}\int_0^{\infty } \frac{1 }{ 1+t^2 \mu^2} \log
   \left(1-\frac{c}{\sqrt{1+t^2}}\right) \,
   dt\right)
  \end{equation}
  and
  \begin{equation}\label{eq:HI}
   H(\mu) = \sqrt{\frac{1+\mu}{1+\nu \mu}} \exp \left( I \right)
  \end{equation}
  where $I$ is the integral of an odd function of $c$,
  \begin{equation}
    I = \int_0^{\infty } \frac{\mu \tanh
   ^{-1}\left(\frac{c}{\sqrt{1+t^2}}\right)}{\pi  \left(1+t^2 \mu^2\right)} \, dt.
  \end{equation}
  This also leads to a new identify for $H$ unique to Flatland,
  \begin{equation}
     H(\mu,c) H(\mu,-c) = \frac{(1+\mu)}{1+\nu \mu}.
  \end{equation}
  Following the same approach and using Eq.(\ref{eq:55}) produces
  \corrected{
  \begin{equation}
    H(\mu) = \exp \left(\int_0^{\infty } \frac{c \, t \, \tan ^{-1}(t \mu)}{\pi  (t^2+1) \left(\sqrt{1+t^2}-c\right)} \, dt\right)
  \end{equation}
  }
  or \remove{the form above} \new{via Eq.(\ref{eq:HI})} with
  \corrected{
  \begin{equation}
    I = \int_0^{\infty } \frac{c \, t \, \tan ^{-1}(t \mu)}{\pi  \sqrt{t^2+1} \left(1-c^2+t^2\right)} \, dt.
  \end{equation}
  }
  Applying the change of variable $y = \sqrt{1+t^2}$, we can write
  \begin{equation}
    \int_0^{\infty } \frac{1 }{ 1+t^2 \mu^2} \log
   \left(1-\frac{c}{\sqrt{1+t^2}}\right) \, dt = \int_{1}^\infty \frac{y \log \left(1-\frac{c}{y}\right)}{\sqrt{y^2-1} \left(\mu^2
   \left(y^2-1\right)+1\right)} dy.
  \end{equation}
  Expanding the integrand
  \begin{equation}
    \frac{y \log \left(1-\frac{c}{y}\right)}{\sqrt{y^2-1} \left(\mu^2
   \left(y^2-1\right)+1\right)} = \sum _{j=1}^{\infty } \left(\frac{c}{y}\right)^j \frac{(-1)^{2 j-1} }{j} \frac{y}{\sqrt{y^2-1} \left(\mu^2
   \left(y^2-1\right)+1\right)}
  \end{equation}
  we note that
  \begin{multline}
    \int_1^\infty \left(\frac{c}{y}\right)^j \frac{(-1)^{2 j-1} }{j} \frac{y}{\sqrt{y^2-1} \left(\mu^2
   \left(y^2-1\right)+1\right)} \, dy = \\ \frac{\sqrt{\pi } (-1)^{2 j+1} c^j \Gamma \left(\frac{j+1}{2}\right) \,
   _2\tilde{F}_1\left(1,\frac{j+1}{2};\frac{j+2}{2};1-\frac{1}{\mu^2}\right)}{2 j \mu^2}
  \end{multline}
  in terms of the regularized Hypergeometric functions $_2\tilde{F}_1$ producing the following series expansions for $I$ over odd powers of $c$,

  \begin{align}
    I &=  \sum_{j(\text{odd}) > 0} \frac{c^j \Gamma \left(\frac{j+1}{2}\right) \,
   _2\tilde{F}_1\left(1,\frac{j+1}{2};\frac{j+2}{2};1-\frac{1}{\mu ^2}\right)}{2
   \sqrt{\pi } j \mu } \\
      &= \sum_{k=0}^\infty \sum_{j(\text{odd}) > 0} \frac{c^j \left(1-\frac{1}{\mu ^2}\right)^k \Gamma
   \left(\frac{j}{2}+k+\frac{1}{2}\right)}{2 \sqrt{\pi } j \mu  \Gamma
   \left(\frac{j}{2}+k+1\right)} \\
      &= \sum_{k=0}^\infty \frac{c \left(1-\frac{1}{\mu ^2}\right)^k \Gamma (k+1) \,
   _3F_2\left(\frac{1}{2},1,k+1;\frac{3}{2},k+\frac{3}{2};c^2\right)}{2 \sqrt{\pi } \mu 
   \Gamma \left(\frac{1}{2} (2 k+3)\right)}.
  \end{align}
  These produce an analytic form of $H$ for the case of normal incidence ($\mu = 1$)
  \begin{equation}
     I_{\mu=1} = \sum_{j(\text{odd}) > 0} \frac{c^j \Gamma \left(\frac{j}{2}+\frac{1}{2}\right)}{2 \sqrt{\pi } j \Gamma \left(\frac{j}{2}+1\right)} = \frac{c \, _3F_2\left(\frac{1}{2},1,1;\frac{3}{2},\frac{3}{2};c^2\right)}{\pi }.
  \end{equation}
  For conservative ($c=1$) scattering the analytic forms
  \begin{multline}
     I_{c=1} = \int_0^{\infty } \frac{\tan ^{-1}(t \mu )}{\pi  \left(\sqrt{1+t^2} t\right)} \, dt \\
     = \frac{2 \left(\sqrt{1-\mu ^2}+1\right) \,
   _3F_2\left(\frac{1}{2},\frac{1}{2},1;\frac{3}{2},\frac{3}{2};-\frac{\left(\sqrt{1-\mu
   ^2}+1\right)^2}{\mu ^2}\right)}{\pi  \mu } \\ +\frac{\left(\log \left(\mu +i \sqrt{1-\mu
   ^2}\right)+\frac{i \pi }{2}\right) \sec ^{-1}(\mu )}{\pi } \\
   = \Re\left(\frac{\, _3F_2\left(\frac{1}{2},1,1;\frac{3}{2},\frac{3}{2};\frac{1}{\mu
   ^2}\right)}{\pi  \mu }\right)
  \end{multline}
  agree with those for normal incidence in the case of normal incident conserative scattering giving
  \begin{equation}
    H(1,c=1) = \sqrt{2} e^{\frac{2 C}{\pi }} = 2.53373727948584190958328963404...
  \end{equation}
  where $C$ is Catalan's constant.  To simplify numerical quadratures it is convenient to move to a finite integration domain.  \corrected{Analogous} to the approach of Stibbs and Weir~\cite{stibbs59} a change of variable $t = \cot \theta$ yields
  \begin{align}\label{eq:Hstibbs}
    H(\mu) &= \frac{(1+\mu) }{1+\nu  \mu} \exp \left(\int_0^{\frac{\pi }{2}} \frac{\mu \log (1+c \sin \theta)}{\pi 
   \left(\mu^2 \cos ^2 \theta+\sin ^2 \theta\right)} \, d \theta\right) \\
   &= \exp \left(\int_0^{\frac{\pi }{2}} \frac{-\mu \log (1-c \sin \theta)}{\pi 
   \left(\mu^2 \cos ^2 \theta+\sin ^2 \theta\right)} \, d \theta\right) \\
   &= \frac{\left((1+\mu) \sqrt{1+c}\right) }{1+\nu  \mu} \exp \left(-\frac{\mu}{\pi } \int_0^{\frac{\pi }{2}} \frac{c
   \cot ^{-1}(\mu \cot \theta) \cos \theta}{\mu+c \mu \sin \theta} \, d\theta\right)
  \end{align}
  or via Eq.(\ref{eq:HI}) with
  \begin{equation}
    I = \int_0^{\frac{\pi }{2}} \frac{\mu \tanh ^{-1}(c \sin \theta)}{\pi  \left(\mu^2 \cos ^2 \theta+\sin
   ^2 \theta\right)} \, d\theta.
  \end{equation}
  We evaluated 38 distinct numerical forms for evaluating $H$ and timed and tested the accuracy of these to 13 significant digits using the NIntegrate function in Mathematica 11.0.0.0 with default parameters.  Additional forms were formed by expanding the integral $I$ in Eq.(\ref{eq:HI}) as a finite sum of powers of $c$ and an integral for the higher powers, for example
\begin{equation}\label{eq:Iexp}
	I = \int_0^{\infty } \frac{c^J t \tan ^{-1}(t \mu)}{\left(1+t^2\right)^{J/2} \left(\pi
   -c^2 \pi +\pi  t^2\right)} \, dt + \sum_{j (\text{odd}) = 1}^{J-2} \frac{c^j \Gamma \left(\frac{1+j}{2}\right) \,
   _2\tilde{F}_1\left(1,\frac{1+j}{2};\frac{2+j}{2};1-\frac{1}{\mu^2}\right)}{2 j \sqrt{\pi
   } \mu}.
\end{equation}
We found Eq.(\ref{eq:Hstibbs}) to be the most efficient form, accurate to 12 significant digits.  The most efficient form accurate to 13 digits was Eqs.(\ref{eq:HI}) and (\ref{eq:Iexp}) with $I$ simplified to
\corrected{
\begin{multline}
   I = \frac{c \left(3 c^4+5 c^2 U+15 U^2\right) \sin ^{-1}\left(\sqrt{U}\right)-c^3
   \left(\left(2 c^2+5\right) \sqrt{1-U} U^{3/2}+3 c^2 \sqrt{(1-U) U}\right)}{15 \pi  \mu
   \sqrt{1-U} U^{5/2}} \\+ \int_0^{\infty } \frac{c^7 t \tan ^{-1}(t \mu)}{\left(1+t^2\right)^{7/2} \left(\pi
   -c^2 \pi +\pi  t^2\right)} \, dt
\end{multline}
}
where $U = 1 - \frac{1}{\mu^2}$.  The median values of all 38 numerical evaluations for $H$, which we take to be likely candidates for the correct results, are sumarized in Table~\ref{table:H}.

\begin{sidewaystable*}[t]
  {\tiny
  \begin{tabular}{c | ccccccccccc}
  & \text{c=0.1} & \text{c=0.5} & \text{c=0.75} & \text{c=0.8} & \text{c=0.9} & \text{c=0.95} & \text{c=0.98} & \text{c=0.99} & \text{c=0.995} & \text{c=0.999} & \text{c=1.0} \\
  \hline \\
 \text{$\mu $=0.1} & 1.009862372199 & 1.057026639875 & 1.097868083873 & 1.108228229927 & 1.133648675540 &
   1.151224135770 & 1.166405350851 & 1.173869060774 & 1.179058103239 & 1.185856006728 &
   1.191238964670 \\
 \text{$\mu $=0.2} & 1.015451175708 & 1.092617888073 & 1.164446809190 & 1.183524834803 & 1.232022663846 &
   1.267122880699 & 1.298631187074 & 1.314575452073 & 1.325853307905 & 1.340884878881 &
   1.353009527684 \\
 \text{$\mu $=0.3} & 1.019554007549 & 1.120238149092 & 1.218958588485 & 1.246080444591 & 1.316905650574 &
   1.369989739920 & 1.419082663070 & 1.444487648046 & 1.462699996597 & 1.487302100642 &
   1.507432578536 \\
 \text{$\mu $=0.4} & 1.022775204139 & 1.142848076656 & 1.265512421387 & 1.300143198575 & 1.392622998417 &
   1.464006928480 & 1.531730871327 & 1.567462480961 & 1.593379507858 & 1.628804711413 &
   1.658159097012 \\
 \text{$\mu $=0.5} & 1.025399949200 & 1.161898371891 & 1.306146347509 & 1.347813653009 & 1.461257535169 &
   1.551107092408 & 1.638314115367 & 1.685136797587 & 1.719463489647 & 1.766895687376 &
   1.806661064765 \\
 \text{$\mu $=0.6} & 1.027592514034 & 1.178259575330 & 1.342115867418 & 1.390389405367 & 1.524084991038 &
   1.632439987778 & 1.739814663753 & 1.798403543768 & 1.841787152039 & 1.902350245534 &
   1.953689405618 \\
 \text{$\mu $=0.7} & 1.029457908725 & 1.192510693704 & 1.374283324352 & 1.428767171444 & 1.581990734535 &
   1.708790148035 & 1.836874033065 & 1.907825045063 & 1.960860751111 & 2.035624288582 &
   2.099679315082 \\
 \text{$\mu $=0.8} & 1.031067806538 & 1.205061773552 & 1.403280536600 & 1.463609138315 & 1.635636873300 &
   1.780740430399 & 1.929953019201 & 2.013791452251 & 2.077026950721 & 2.167010237191 &
   2.244904668164 \\
 \text{$\mu $=0.9} & 1.032473415474 & 1.216215837575 & 1.429590443702 & 1.495425861272 & 1.685543022198 &
   1.848748894839 & 2.019405386361 & 2.116593222028 & 2.190532595386 & 2.296707939516 &
   2.389548190777 \\
 \text{$\mu $=1.0} & 1.033712591466 & 1.226203880894 & 1.453593356088 & 1.524622994060 & 1.732130621233 &
   1.913189719542 & 2.105516087365 & 2.216458389328 & 2.301565481862 & 2.424860908147 &
   2.533737279485 \\
\end{tabular}}
\caption{Benchmark values for the Flatland $H$-function $H(\mu)$.\label{table:H}}
\end{sidewaystable*} 
\bibliographystyle{elsarticle-num-names}
\bibliography{flatlandttsp}

\begin{thebibliography}{27}
\providecommand{\natexlab}[1]{#1}
\providecommand{\url}[1]{\texttt{#1}}
\providecommand{\urlprefix}{URL }
\expandafter\ifx\csname urlstyle\endcsname\relax
  \providecommand{\doi}[1]{doi:\discretionary{}{}{}#1}\else
  \providecommand{\doi}[1]{doi:\discretionary{}{}{}\begingroup
  \urlstyle{rm}\url{#1}\endgroup}\fi
\providecommand{\bibinfo}[2]{#2}

\bibitem[{Davison(1957)}]{davison57}
\bibinfo{author}{B.~Davison}, \bibinfo{title}{{Neutron Transport Theory}},
  \bibinfo{publisher}{Oxford University Press}, \bibinfo{year}{1957}.

\bibitem[{Chandrasekhar(1960)}]{chandrasekhar60}
\bibinfo{author}{S.~Chandrasekhar}, \bibinfo{title}{Radiative Transfer},
  \bibinfo{publisher}{Dover}, \bibinfo{year}{1960}.

\bibitem[{Wing(1962)}]{wing62}
\bibinfo{author}{G.~Wing}, \bibinfo{title}{An introduction to transport
  theory}, \bibinfo{publisher}{Wiley}, \bibinfo{year}{1962}.

\bibitem[{Hoogenboom(2008)}]{hoogenboom08a}
\bibinfo{author}{J.~Hoogenboom}, \bibinfo{title}{The Two-Direction
  Neutral-Particle Transport Model: A Useful Tool for Research and Education},
  \bibinfo{journal}{Transport Theory and Statistical Physics}
  \bibinfo{volume}{37}~(\bibinfo{number}{1}) (\bibinfo{year}{2008})
  \bibinfo{pages}{65--108}.

\bibitem[{Adams et~al.(1989)Adams, Larsen, and Pomraning}]{adams89}
\bibinfo{author}{M.~Adams}, \bibinfo{author}{E.~Larsen},
  \bibinfo{author}{G.~Pomraning}, \bibinfo{title}{Benchmark results for
  particle transport in a binary Markov statistical medium},
  \bibinfo{journal}{Journal of Quantitative Spectroscopy and Radiative
  Transfer} \bibinfo{volume}{42}~(\bibinfo{number}{4}) (\bibinfo{year}{1989})
  \bibinfo{pages}{253--266}.

\bibitem[{Paasschens(1997)}]{paasschens97}
\bibinfo{author}{J.~C.~J. Paasschens}, \bibinfo{title}{Solution of the
  time-dependent Boltzmann equation}, \bibinfo{journal}{Phys. Rev. E}
  \bibinfo{volume}{56}~(\bibinfo{number}{1}) (\bibinfo{year}{1997})
  \bibinfo{pages}{1135--1141}, \doi{\bibinfo{doi}{10.1103/PhysRevE.56.1135}}.

\bibitem[{Jarosz et~al.(2012)Jarosz, Sch{\"o}nefeld, Kobbelt, and
  Jensen}]{jarosz12}
\bibinfo{author}{W.~Jarosz}, \bibinfo{author}{V.~Sch{\"o}nefeld},
  \bibinfo{author}{L.~Kobbelt}, \bibinfo{author}{H.~W. Jensen},
  \bibinfo{title}{Theory, analysis and applications of 2D global illumination},
  \bibinfo{journal}{ACM Transactions on Graphics (TOG)}
  \bibinfo{volume}{31}~(\bibinfo{number}{5}) (\bibinfo{year}{2012})
  \bibinfo{pages}{125}.

\bibitem[{Yang and Deb(2009)}]{yang09}
\bibinfo{author}{X.-S. Yang}, \bibinfo{author}{S.~Deb}, \bibinfo{title}{Cuckoo
  search via L{\'e}vy flights}, in: \bibinfo{booktitle}{Nature \& Biologically
  Inspired Computing, 2009. NaBIC 2009. World Congress on},
  \bibinfo{organization}{IEEE}, \bibinfo{pages}{210--214},
  \bibinfo{year}{2009}.

\bibitem[{Bal et~al.(2000)Bal, Freilikher, Papanicolaou, and
  Ryzhik}]{bal2000wave}
\bibinfo{author}{G.~Bal}, \bibinfo{author}{V.~Freilikher},
  \bibinfo{author}{G.~Papanicolaou}, \bibinfo{author}{L.~Ryzhik},
  \bibinfo{title}{Wave transport along surfaces with random impedance},
  \bibinfo{journal}{Physical Review B}
  \bibinfo{volume}{62}~(\bibinfo{number}{10}) (\bibinfo{year}{2000})
  \bibinfo{pages}{6228}.

\bibitem[{Meylan and Masson(2006)}]{meylan06}
\bibinfo{author}{M.~H. Meylan}, \bibinfo{author}{D.~Masson}, \bibinfo{title}{A
  linear Boltzmann equation to model wave scattering in the marginal ice zone},
  \bibinfo{journal}{Ocean Modelling} \bibinfo{volume}{11}~(\bibinfo{number}{3})
  (\bibinfo{year}{2006}) \bibinfo{pages}{417--427}.

\bibitem[{Vynck et~al.(2012)Vynck, Burresi, Riboli, and Wiersma}]{vynck12}
\bibinfo{author}{K.~Vynck}, \bibinfo{author}{M.~Burresi},
  \bibinfo{author}{F.~Riboli}, \bibinfo{author}{D.~S. Wiersma},
  \bibinfo{title}{Photon management in two-dimensional disordered media},
  \bibinfo{journal}{Nature materials}
  \bibinfo{volume}{11}~(\bibinfo{number}{12}) (\bibinfo{year}{2012})
  \bibinfo{pages}{1017--1022}.

\bibitem[{Pearson(1905)}]{pearson05}
\bibinfo{author}{K.~Pearson}, \bibinfo{title}{The problem of the random walk},
  \bibinfo{journal}{Nature} \bibinfo{volume}{72}~(\bibinfo{number}{1865})
  (\bibinfo{year}{1905}) \bibinfo{pages}{294}.

\bibitem[{Kluyver(1906)}]{kluyver06}
\bibinfo{author}{J.~Kluyver}, \bibinfo{title}{A local probability problem},
  \bibinfo{journal}{Nederl. Acad. Wetensch. Proc} \bibinfo{volume}{8}
  (\bibinfo{year}{1906}) \bibinfo{pages}{341--350}.

\bibitem[{Rayleigh(1919)}]{rayleigh19}
\bibinfo{author}{L.~Rayleigh}, \bibinfo{title}{XXXI. On the problem of random
  vibrations, and of random flights in one, two, or three dimensions},
  \bibinfo{journal}{The London, Edinburgh, and Dublin Philosophical Magazine
  and Journal of Science} \bibinfo{volume}{37}~(\bibinfo{number}{220})
  (\bibinfo{year}{1919}) \bibinfo{pages}{321--347}.

\bibitem[{Grosjean(1953)}]{grosjean53}
\bibinfo{author}{C.~C. Grosjean}, \bibinfo{title}{{Solution of the
  non-isotropic random flight problem in the k-dimensional space}},
  \bibinfo{journal}{Physica} \bibinfo{volume}{19}~(\bibinfo{number}{1-12})
  (\bibinfo{year}{1953}) \bibinfo{pages}{29--45}, ISSN
  \bibinfo{issn}{0031-8914}.

\bibitem[{Watson(1962)}]{watson62}
\bibinfo{author}{G.~N. Watson}, \bibinfo{title}{A Treatise on the Theory of
  Bessel Functions}, \bibinfo{publisher}{Cambridge University Press},
  \bibinfo{edition}{2nd} edn., \bibinfo{year}{1962}.

\bibitem[{Liemert and Kienle(2011)}]{liemert11c}
\bibinfo{author}{A.~Liemert}, \bibinfo{author}{A.~Kienle},
  \bibinfo{title}{Radiative transfer in two-dimensional infinitely extended
  scattering media}, \bibinfo{journal}{Journal of Physics A: Mathematical and
  Theoretical} \bibinfo{volume}{44} (\bibinfo{year}{2011})
  \bibinfo{pages}{505206}.

\bibitem[{Zoia et~al.(2011)Zoia, Dumonteil, and Mazzolo}]{zoia11e}
\bibinfo{author}{A.~Zoia}, \bibinfo{author}{E.~Dumonteil},
  \bibinfo{author}{A.~Mazzolo}, \bibinfo{title}{Collision densities and mean
  residence times for d-dimensional exponential flights},
  \bibinfo{journal}{Physical Review E}
  \bibinfo{volume}{83}~(\bibinfo{number}{4}) (\bibinfo{year}{2011})
  \bibinfo{pages}{041137}.

\bibitem[{d'Eon(2014)}]{deon14}
\bibinfo{author}{E.~d'Eon}, \bibinfo{title}{Rigorous Asymptotic and
  Moment-Preserving Diffusion Approximations for Generalized Linear Boltzmann
  Transport in Arbitrary Dimension}, \bibinfo{journal}{Transport Theory and
  Statistical Physics} \bibinfo{volume}{42}~(\bibinfo{number}{6-7})
  (\bibinfo{year}{2014}) \bibinfo{pages}{237--297},
  \doi{\bibinfo{doi}{10.1080/00411450.2014.910231}},
  \urlprefix\url{http://dx.doi.org/10.1080/00411450.2014.910231}.

\bibitem[{Asadzadeh and Larsen(2008)}]{asadzadeh08}
\bibinfo{author}{M.~Asadzadeh}, \bibinfo{author}{E.~Larsen},
  \bibinfo{title}{Linear transport equations in flatland with small angular
  diffusion and their finite element approximations},
  \bibinfo{journal}{Mathematical and Computer Modelling}
  \bibinfo{volume}{47}~(\bibinfo{number}{3-4}) (\bibinfo{year}{2008})
  \bibinfo{pages}{495--514}.

\bibitem[{Liemert and Kienle(2012{\natexlab{a}})}]{liemert12c}
\bibinfo{author}{A.~Liemert}, \bibinfo{author}{A.~Kienle},
  \bibinfo{title}{Green's functions for the two-dimensional radiative transfer
  equation in bounded media}, \bibinfo{journal}{Journal of Physics A:
  Mathematical and Theoretical} \bibinfo{volume}{45}~(\bibinfo{number}{17})
  (\bibinfo{year}{2012}{\natexlab{a}}) \bibinfo{pages}{175201--175209}.

\bibitem[{Liemert and Kienle(2012{\natexlab{b}})}]{liemert12a}
\bibinfo{author}{A.~Liemert}, \bibinfo{author}{A.~Kienle},
  \bibinfo{title}{Analytical approach for solving the radiative transfer
  equation in two-dimensional layered media}, \bibinfo{journal}{Journal of
  Quantitative Spectroscopy and Radiative Transfer}
  \bibinfo{volume}{113}~(\bibinfo{number}{7})
  (\bibinfo{year}{2012}{\natexlab{b}}) \bibinfo{pages}{559--564}.

\bibitem[{Machida(2015)}]{machida15}
\bibinfo{author}{M.~Machida}, \bibinfo{title}{The radiative transport equation
  in flatland with separation of variables}, \bibinfo{journal}{arXiv preprint
  arXiv:1511.05723} .

\bibitem[{Stewart et~al.(1966)Stewart, Ku{\v{s}}{\v{c}}er, and
  McCormick}]{stewart1966equivalence}
\bibinfo{author}{J.~C. Stewart}, \bibinfo{author}{I.~Ku{\v{s}}{\v{c}}er},
  \bibinfo{author}{N.~J. McCormick}, \bibinfo{title}{Equivalence of special
  models in energy-dependent neutron transport and nongrey radiative transfer},
  \bibinfo{journal}{Annals of Physics}
  \bibinfo{volume}{40}~(\bibinfo{number}{2}) (\bibinfo{year}{1966})
  \bibinfo{pages}{321--333}.

\bibitem[{Williams(1971)}]{williams71}
\bibinfo{author}{M.~M.~R. Williams}, \bibinfo{title}{Mathematical methods in
  particle transport theory.}, \bibinfo{publisher}{Wiley},
  \bibinfo{year}{1971}.

\bibitem[{Placzek and Seidel(1947)}]{placzek47b}
\bibinfo{author}{G.~Placzek}, \bibinfo{author}{W.~Seidel},
  \bibinfo{title}{Milne's Problem in Transport Theory}, \bibinfo{journal}{Phys.
  Rev.} \bibinfo{volume}{72}~(\bibinfo{number}{7}) (\bibinfo{year}{1947})
  \bibinfo{pages}{550--555}, \doi{\bibinfo{doi}{10.1103/PhysRev.72.550}}.

\bibitem[{Stibbs and Weir(1959)}]{stibbs59}
\bibinfo{author}{D.~Stibbs}, \bibinfo{author}{R.~Weir}, \bibinfo{title}{On the
  {H}-functions for isotropic scattering}, \bibinfo{journal}{Monthly Notices of
  the Royal Astronomical Society} \bibinfo{volume}{119} (\bibinfo{year}{1959})
  \bibinfo{pages}{512}.

\end{thebibliography}

\end{document}